\documentclass[twocolumn,secnumarabic,amssymb, nobibnotes, aps, prl, superscriptaddress]{revtex4-2}

\newcommand*\vtick{\textsc{\char13}}
\usepackage[]{graphicx}
\usepackage{tabularx}
\usepackage[usenames,dvipsnames]{color}
\usepackage{soul}
\usepackage[normalem]{ulem}
\usepackage{bm}
\usepackage{amsmath}
\usepackage{lineno}
\usepackage{gensymb}
\usepackage{tabularx}
\usepackage{amsfonts}
\usepackage{mathrsfs}
\usepackage{graphicx}% Include figure files
\usepackage{dcolumn}% Align table columns on decimal point
\usepackage{bm}% bold math
\usepackage{color}
\usepackage[colorlinks,bookmarks=false,citecolor=blue,linkcolor=red,urlcolor=blue]{hyperref}
\usepackage{multirow}
\usepackage{physics}
\usepackage{multirow}


\begin{document}

\title{Molecule-intercalated van der Waals Josephson junctions uncover complex multiband superconductivity}

\author{Daniel Margineda}
\thanks{Corresponding author}
\email{d.margineda@gmail.com}
\affiliation{CIC nanoGUNE BRTA, 20018, San Sebastián, Spain}
%\email{d.margineda@nanogune.eu}
%\email{d.margineda@gmail.com}
\author{David Caldevilla-Asenjo}
\affiliation{Centro de Física de Materiales – Materials Physics Center (CFM-MPC) CSIC-EHU, 20018, San Sebastian, Spain}

\author{ Covadonga  \'Alvarez-Garc\'ia}
\affiliation{CIC nanoGUNE BRTA, 20018, San Sebastián, Spain}

\affiliation{Departamento de Polímeros y Materiales Avanzados: Física, Química y Tecnología, Facultad de Químicas (EHU), 20018, San Sebastián, Spain} 

\author{ Daniel Tezze}
\affiliation{CIC nanoGUNE BRTA, 20018, San Sebastián, Spain}

\affiliation{Departamento de Polímeros y Materiales Avanzados: Física, Química y Tecnología, Facultad de Químicas (EHU), 20018, San Sebastián, Spain}

\author{Raul Arenal}
\affiliation{Instituto de Nanociencia y Materiales de Aragón (INMA) CSIC-Universidad de Zaragoza, 50009, Zaragoza, Spain}
\affiliation{Laboratorio de Microscopías Avanzadas (LMA), Universidad de Zaragoza, Zaragoza, Spain}
\affiliation{ARAID Foundation, 50018 Zaragoza, Spain}

\author{Maxim Ilyn}
\affiliation{Centro de Física de Materiales – Materials Physics Center (CFM-MPC) CSIC-EHU, 20018, San Sebastian, Spain}
\author{Celia Rogero}
\affiliation{Centro de Física de Materiales – Materials Physics Center (CFM-MPC) CSIC-EHU, 20018, San Sebastian, Spain}
\author{Luis E. Hueso}
\affiliation{CIC nanoGUNE BRTA, 20018, San Sebastián, Spain}
\affiliation{Ikerbasque, Basque Foundation for Science, 48009, Bilbao, Spain}
\author{Marco Gobbi}
\affiliation{Centro de Física de Materiales – Materials Physics Center (CFM-MPC) CSIC-EHU, 20018, San Sebastian, Spain}
\affiliation{Ikerbasque, Basque Foundation for Science, 48009, Bilbao, Spain}
\begin{abstract}
Intercalation of molecular species provides a route to tailor the electronic structure in van der Waals superconductors, yet intercalated compounds have rarely been integrated into phase-sensitive quantum devices. Here, we realize hybrid molecular Josephson junctions based on TaS$_2$/NbSe$_2$ heterostructures. Selective intercalation of the TaS$_2$ electrode induces monolayer-like Ising superconductivity, while the junctions exhibit asymmetric Fraunhofer interference, field-induced and spontaneous Josephson diode effects. The zero-field diode disappears upon reversing the electrode stacking and re-emerges after intercalating the lower NbSe$_2$ electrode. These findings indicate that the hybrid superconducting electrode hosts unconventional superconductivity underlying the nonreciprocal transport, while molecular intercalation acts as a tuning knob for its internal phase structure. This behavior is consistent with a time-reversal-symmetry-breaking multiband superconducting state characterized by a nontrivial relative phase between superconducting condensates. Our results establish hybrid molecular Josephson junctions as a quantum-device platform for phase-sensitive investigation and integration of molecularly engineered superconductors, opening opportunities to uncover and engineer unconventional superconducting phases.

\end{abstract}
\keywords{van der Waals superconductors, Ising superconductivity, organic intercalation, Time-reversal symmetry breaking, Multiband superconductivity, Josephson diode effect, Nonreciprocal transport, S+is' pairing state}

\maketitle

\begin{figure*}[ht]
\centering
\includegraphics[scale=0.26 ]{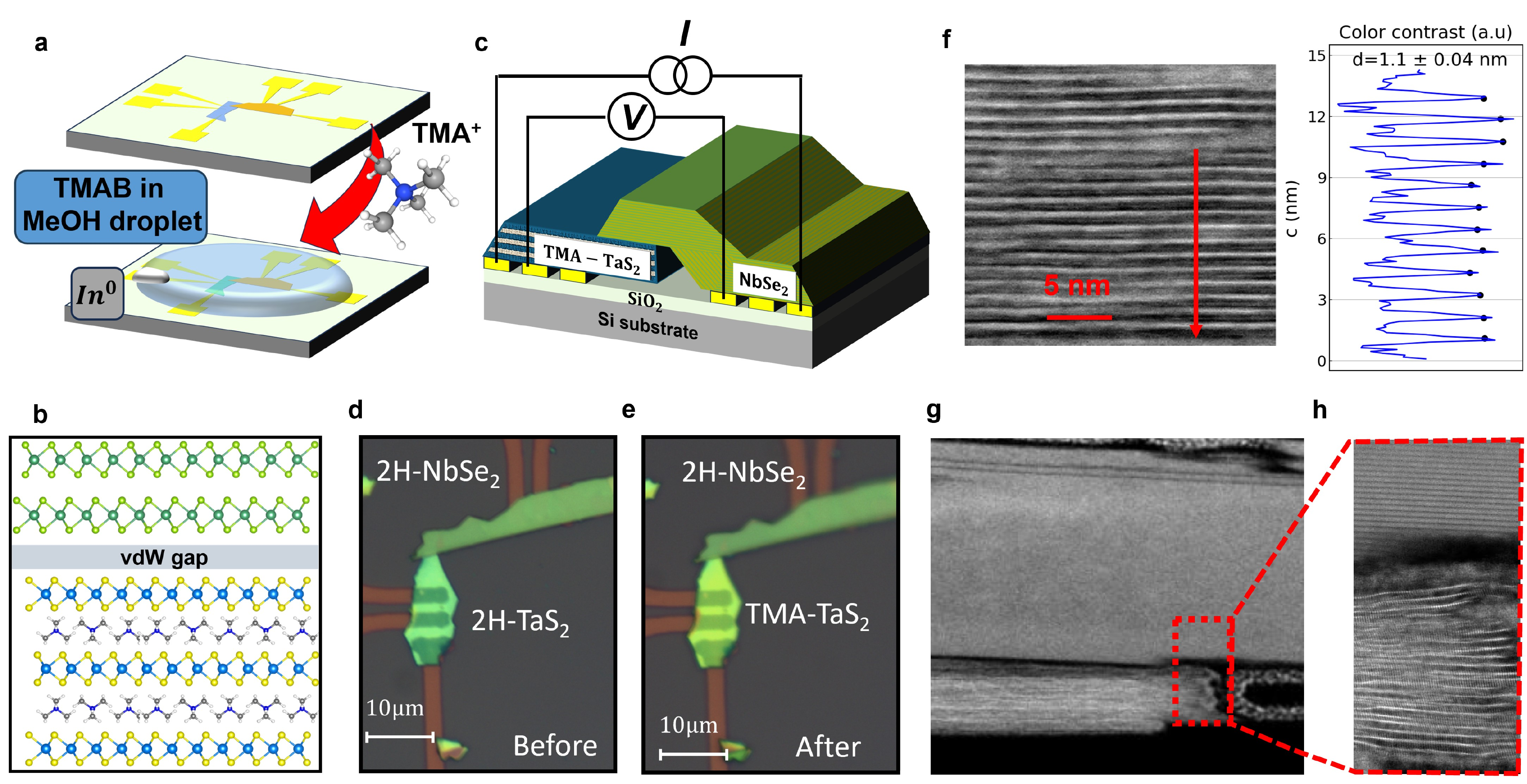}
\caption{
\textbf{Hybrid molecular van der Waals Josephson heterojunction.}
\textbf{a,} Schematic of the selective galvanic intercalation process. An indium sacrificial anode electrically connected to a TaS$_2$/NbSe$_2$ heterostructure drives the insertion of tetramethylammonium (TMA$^+$) cations into the TaS$_2$ crystal. 
\textbf{b,} Concept of the hybrid molecular Josephson junction consisting of a molecularly intercalated TaS$_2$ electrode vertically coupled to a pristine 2H-NbSe$_2$ superconductor through a vdW interface. \textbf{c,} Device architecture and measurement configuration. Independent bottom Au contacts provide separate electrical access to the hybrid superconducting electrode and the Josephson junction. \textbf{d,e}  Optical micrographs of the heterostructure before and after intercalation, respectively. The homogeneous optical contrast change is observed exclusively in the TaS$_2$ crystal. \textbf{f,} Cross-sectional STEM-HAADF image of the intercalated TaS$_2$ crystal and the corresponding intensity line profile, revealing the expanded interlayer spacing induced by molecular intercalation. \textbf{g,} Cross-sectional STEM-HAADF image of a high-resistance TaS$_2$/NbSe$_2$ heterojunction. \textbf{h,} Magnified STEM-HAADF image of the junction interface showing local structural disorder and an interfacial oxide layer.}
\label{fig1} 
\end{figure*}

\noindent\large{\textbf{Introduction}}\normalsize\\ 
Identifying, controlling and integrating unconventional superconducting states into functional quantum devices remain key hurdles. Transition-metal dichalcogenides (TMDs) host intertwined multiband electronic structures, strong spin–orbit coupling and competing superconducting and charge-density-wave orders~\cite{Manzeli2017,Maj2020}, providing fertile ground for unconventional superconductivity~\cite{Lane22}. In the prototypical TMD superconductor TaS$_2$, deviations from conventional isotropic s-wave superconductivity have been reported in the two-dimensional limit ~\cite{1HTaS2_unc,gal14,mar26} and in mixed-stacking 1H/1T polytypes~\cite{chiral4Hb,Sazgari2026G}.

Molecular intercalation offers a further, highly tunable strategy for engineering superconducting states by modifying interlayer coupling and carrier density~\cite{Mar2025,Li21,zhang2022}. Signatures of unconventional superconductivity have been reported in TaS$_2$ intercalated with chiral molecules~\cite{24_chiral}. Whether these apparently distinct manifestations share a common microscopic origin, and how molecular intercalation influences the superconducting symmetry and internal phase structure, remain unclear~\cite{Mar2025,pi_shift}.

Addressing these questions requires phase-sensitive probes capable of accessing the superconducting order parameter and its underlying symmetries. The superconducting diode effect (SDE), manifested as unequal critical currents for opposite current directions~\cite{and20,mar25}, is a powerful probe of symmetry breaking~\cite{nad23}. While field-induced nonreciprocity arises from mechanisms that lift inversion symmetry~\cite{bau22,lyu21A,mar23,hou23,cas25,Fukaya2025}, superconducting states with internal phase degrees of freedom that spontaneously break time-reversal symmetry can generate nonreciprocal transport even at zero magnetic field~\cite{zhao23,Le2024,Anwar2023,Qiu2023}.

Van der Waals (vdW) Josephson junctions provide a natural platform for phase-sensitive investigation of layered superconductors~\cite{Yabuki2016,Ai2021,Idzuchi2021,Lee2021,HuG2023}. However, molecule-intercalated superconductors have so far been investigated primarily through material characterization and conventional transport, while their integration into Josephson devices is essentially unexplored. This experimental gap limits phase-sensitive access to their superconducting states and their development for quantum-device applications.

Here, we introduce hybrid molecular Josephson junctions based on molecularly engineered van der Waals superconductors. Starting from TaS$_2$/NbSe$_2$ heterostructures, we selectively intercalate TMA ions into the TaS$_2$ flake, driving it into a monolayer-like Ising superconducting regime. The resulting TMA-TaS$_2$/NbSe$_2$ junction retains coherent Josephson transport and exhibits an asymmetric Fraunhofer interference pattern indicative of inversion-symmetry breaking, together with spontaneous and field-induced nonreciprocal transport. The zero-field diode response disappears when the stacking order of TMA-TaS$_2$ and NbSe$_2$ is reversed but emerges again after intercalation of the lower NbSe$_2$ electrode. This electrode-selective behavior disfavors a purely interfacial origin and instead links the spontaneous response with the intercalation-modified superconducting state of the lower electrode. Our results are consistent with the emergence of a complex multiband s+is\vtick\ superconducting state, arising from the interplay of competing superconducting states and interband scattering that may induce a nontrivial relative phase, spontaneously breaking time-reversal symmetry.\\

\noindent\large{\textbf{Results}}\normalsize\\
We fabricated hybrid vdW Josephson junctions consisting of a molecularly intercalated TaS$_2$ electrode vertically coupled to a pristine 2H-NbSe$_2$ superconductor. The heterostructures were assembled by polymer-assisted dry-transfer technique under inert atmosphere.
Molecular intercalation of tetramethylammonium (TMA) cations was subsequently achieved using our recently developed galvanic approach~\cite{Tezze2026,Mar2025}. By selecting indium as a sacrificial anode with a reduction potential between the intercalation potentials of TaS$_2$ and NbSe$_2$, molecular intercalation is selectively confined to the TaS$_2$ crystal, while preserving the pristine NbSe$_2$ electrode (Fig.~\ref{fig1}a,b). The device architecture enables independent characterization of the molecularly engineered superconductor and the coherent Josephson transport across the electrode interface (Fig.~\ref{fig1}c).

The resulting TMA-TaS$_2$ superlattices exhibit a uniform optical contrast change restricted to the TaS$_2$ flake (Fig.~\ref{fig1}d,e), indicating homogeneous molecular intercalation. Scanning transmission electron microscopy high-angle dark field (STEM-HAADF) reveals highly ordered stacking with an increased interlayer periodicity of $d_{int} \simeq$ 1.1\,nm (Fig.~\ref{fig1}f and Extended Data Fig.\,1).

\begin{figure*}[ht]
\centering
\includegraphics[scale=0.27 ]{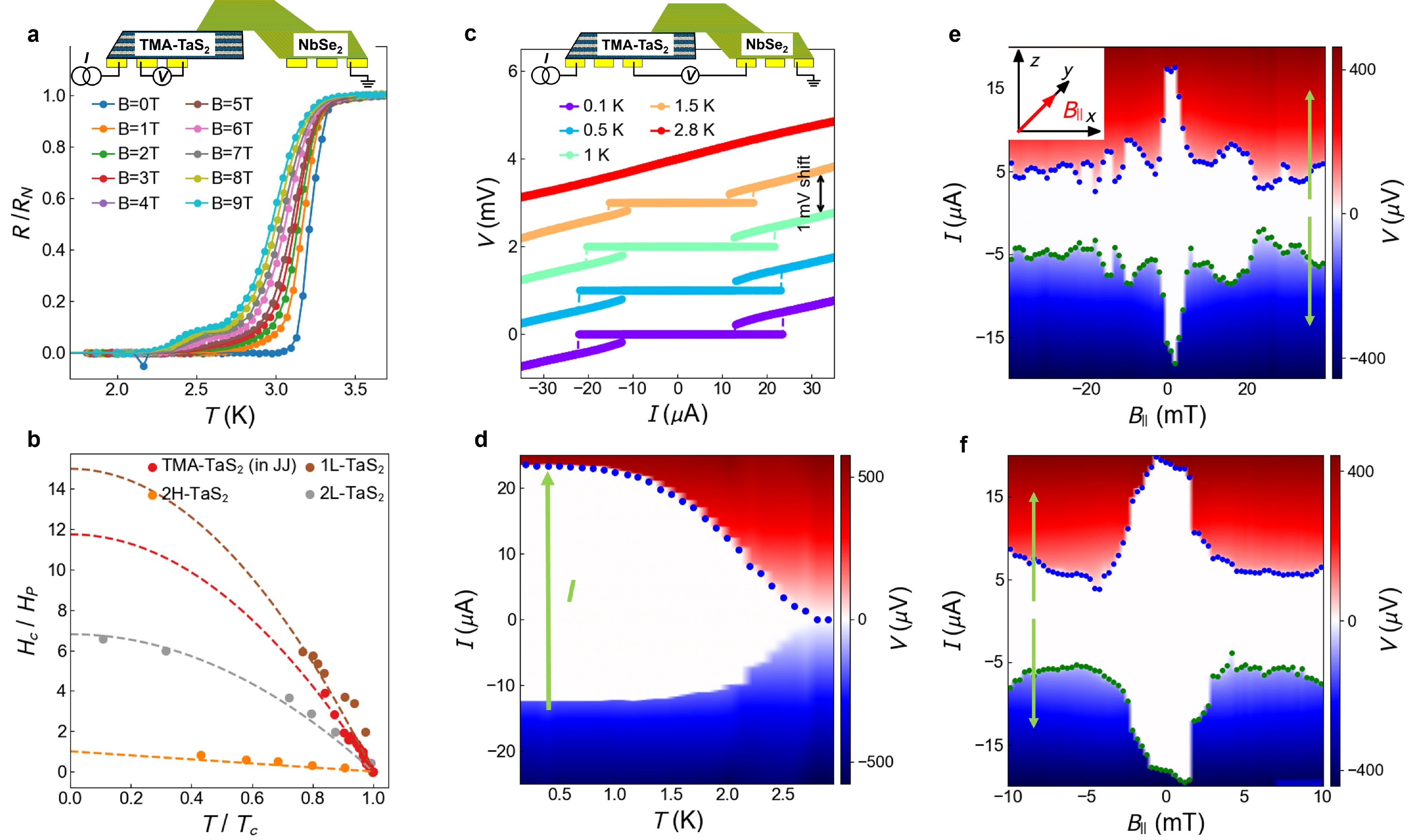}
\caption{
\textbf{Monolayer-like Ising superconductivity and Josephson coupling in hybrid molecular heterojunctions.}
\textbf{a,} Temperature dependence of the normalized resistance of the molecular TaS$_2$ electrode measured under different in-plane magnetic fields.
\textbf{b,} Reduced in-plane critical field ($H_{\mathrm c}/H_{\mathrm P}$) as a function of reduced temperature ($T/T_c$) for molecularly intercalated TaS$_2$, compared with reported values for monolayer, bilayer and bulk TaS$_2$.
\textbf{c,} Current-voltage characteristics of the TMA-TaS$_2$/NbSe$_2$ Josephson junction measured at different temperatures, showing hysteretic switching of an underdamped junction.
\textbf{d,} Temperature dependence of the switching current of the hybrid Josephson junction, demonstrating Josephson coupling above the superconducting transition temperature of pristine TaS$_2$, consistent with the enhanced superconductivity of the intercalated electrode.
\textbf{e,} Asymmetric Fraunhofer interference pattern measured under in-plane magnetic field showing multiple interference lobes.
\textbf{f,} Low-field, high-resolution measurement of the Fraunhofer interference pattern after zero-field cooling. Nonreciprocal transport together with a maximum switching current at finite field is observed.}
\label{fig2} 
\end{figure*}
STEM and energy-dispersive X-ray spectroscopy further confirm selective intercalation of TaS$_2$, while the NbSe$_2$ electrode remains unaffected (Fig.~\ref{fig1}g,h and Extended Data Fig.\,2).

\begin{figure*}[ht]
\centering
\includegraphics[scale=0.29 ]{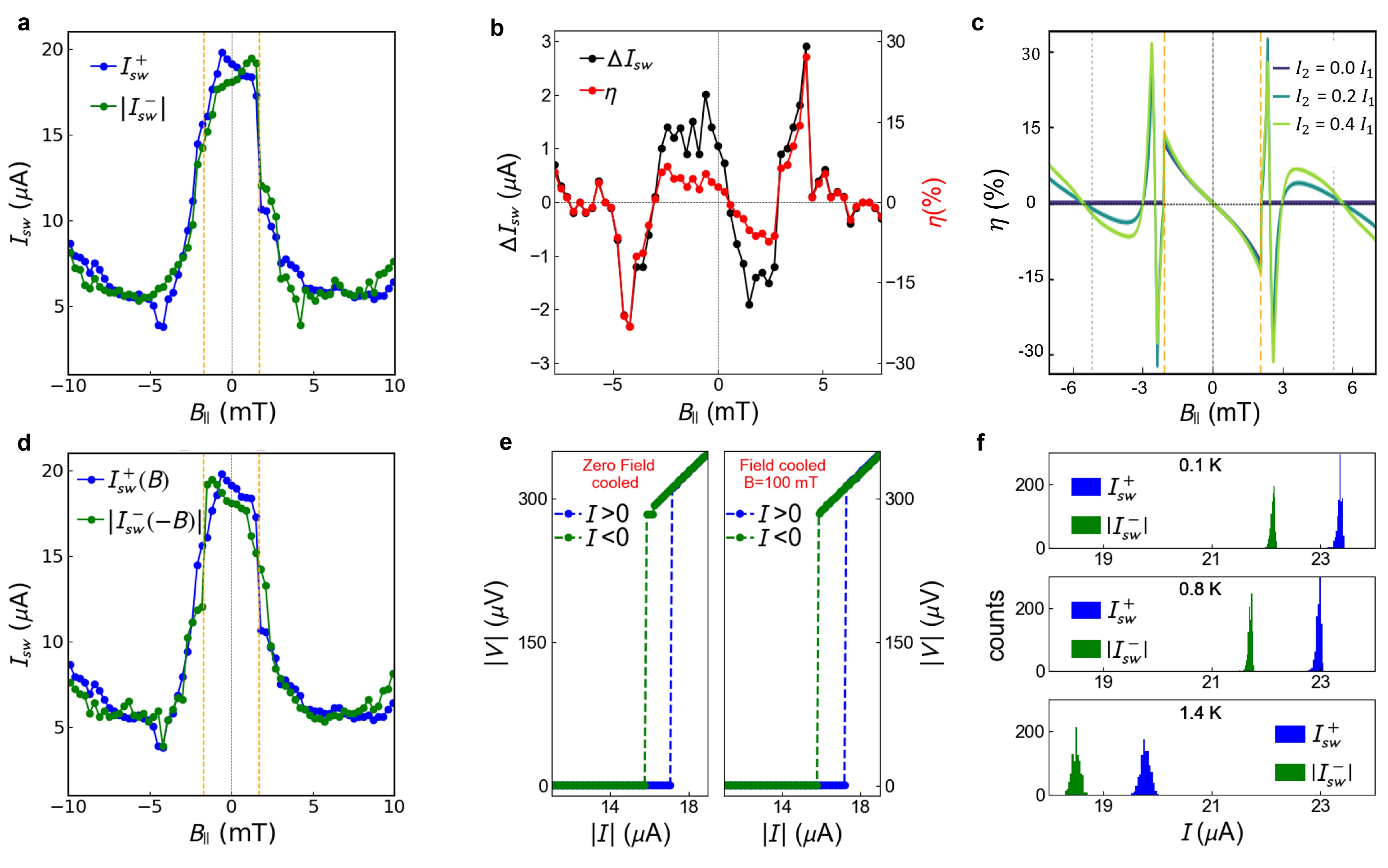}
\caption{\textbf{Field-induced and spontaneous Josephson diode effect.}
\textbf{a,} Positive ($I_{\rm sw}^{+}$) and negative ($|I_{\rm sw}^{-}|$) switching currents measured as a function of in-plane magnetic field. The switching currents exhibit maxima at a finite field ($B\simeq B^\star$, orange line).
\textbf{b,} Switching-current difference ($\Delta I_{\rm sw}$) and diode efficiency ($\eta$) as a function of magnetic field,  revealing zero-field nonreciprocity together with a field-induced sign reversal.
\textbf{c,} Phenomenological modelling of the field-induced nonreciprocal response (Methods).
Calculated diode efficiency as a function of magnetic field for different second-harmonic amplitudes at fixed vortex position $(x_0,z_0)=(0.1W,0.1W)$ and phase winding $\lambda_v=1$. \textbf{d,} Symmetry analysis of the field-dependent switching currents comparing $I_{\rm sw}^{+}(B)$ and $|I_{\rm sw}^{-}(-B)|$, revealing an even-in-field component below $B^\star$. \textbf{e,} Current-voltage characteristics measured after zero-field cooling (left) and field cooling at $B=100$\,mT (right). The magnitude and sign of the spontaneous nonreciprocal response remain unchanged within the experimental uncertainty.
\textbf{f,} Switching-current histograms measured at different temperatures. The finite separation between the positive ($I_{\rm sw}^{+}$) and negative ($|I_{\rm sw}^{-}|$) switching-current distributions persists throughout the investigated temperature range.}
\label{fig3} 
\end{figure*}

We first investigate the superconducting properties of the hybrid TaS$_2$ electrode. The temperature dependence of the resistance under in-plane magnetic field reveals an enhanced critical temperature and strong resilience against Zeeman depairing, indicative of monolayer Ising superconductivity (Fig.~\ref{fig2}a). To quantify the Ising enhancement, we compare the in-plane critical field with reported values for single-layer and bilayer TaS$_2$~\cite{yang_enhanced_2018}. The critical fields substantially exceed the Pauli limit, $H_P=1.85 \,k_B T_c$, and approach those reported in the monolayer limit (Fig.~\ref{fig2}b). The Ising enhancement demonstrates effective electronic interlayer decoupling within the molecular superlattice and serves as a sensitive indicator of intercalation quality~\cite{Mar2025}.

We next investigate the Josephson response of the hybrid heterojunctions by measuring the voltage drop across the two superconducting electrodes (Fig.~\ref{fig2}c,d). The current-voltage characteristics exhibit hysteretic switching between the superconducting and resistive states, consistent with an underdamped Josephson junction. The switching current remains finite up to T$\simeq$ 2.8\,K, well above the superconducting transition temperature of pristine TaS$_2$ ($\sim$ 1\,K), demonstrating that molecular intercalation enhances superconductivity while preserving coherent Josephson transport.

The junction exhibits a Fraunhofer-like magnetic-field interference pattern with multiple lobes, finite minima and pronounced lobe asymmetries (Fig.~\ref{fig2}e). Nonzero minima are commonly associated with nonuniform critical-current distributions~\cite{barone82} and are frequently observed in van der Waals Josephson junctions~\cite{Ai2021,Idzuchi2021,Kang2022,dvir21}. The asymmetric lobes are consistent with a nonuniform phase distribution arising from a spatially inhomogeneous magnetic environment~\cite{kras20,clem11,Mar17}. In addition, because the micrometre-scale lateral dimensions exceed the Pearl length, screening currents substantially influence the low-field response.

\begin{figure*}[ht]
\centering
\includegraphics[scale=0.26 ]{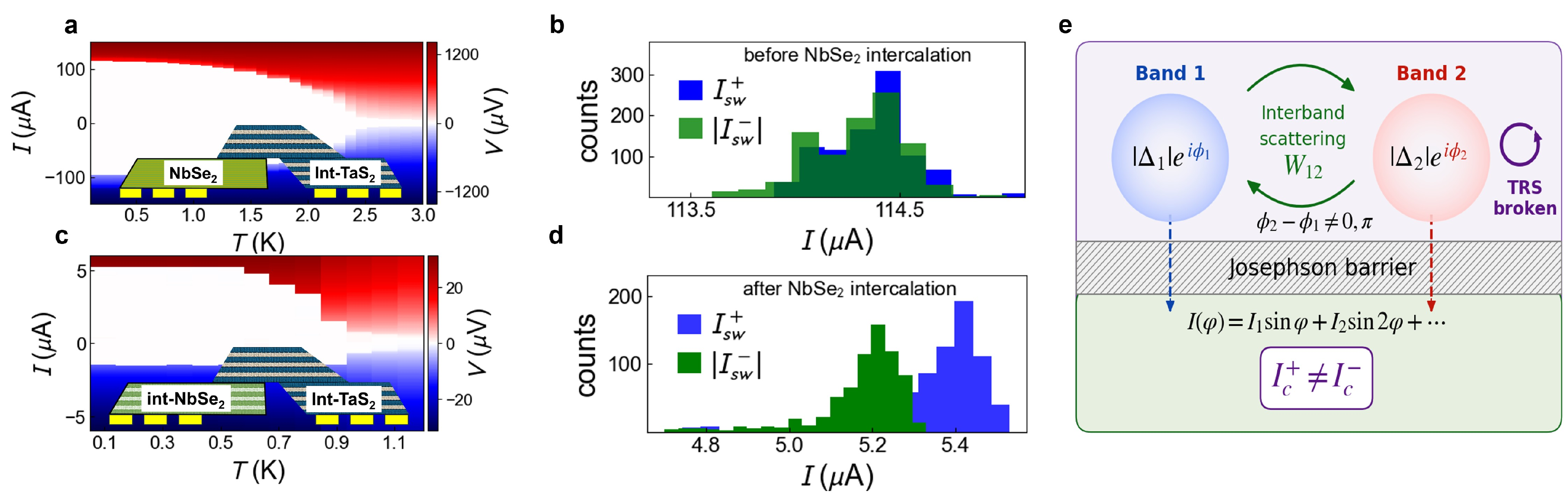}
\caption{
\textbf{Reversed heterostructure, NbSe$_2$ intercalation and microscopic interpretation of the spontaneous Josephson diode effect.}
\textbf{a,}
Temperature dependence of the current-voltage characteristics of the reversed NbSe$_2$/TMA-TaS$_2$ Josephson junction, showing superconducting behavior comparable to the original device.
\textbf{b,}
Switching-current distributions of the reversed heterostructure measured at $T=100$ mK, showing no detectable zero-field Josephson diode response.
\textbf{c,}
Temperature dependence of the current-voltage characteristics after CTA$^+$ intercalation of the lower NbSe$_2$ electrode. The superconducting transition shifts to $T_c\simeq0.9$ K, consistent with monolayer-like superconductivity in NbSe$_2$.
\textbf{d,}
Switching-current distributions after NbSe$_2$ intercalation, revealing the recovery of the zero-field Josephson diode response.
\textbf{e,} Proposed microscopic scenario for the spontaneous Josephson diode effect. Interband scattering between superconducting condensates with similar energies stabilizes an anomalous relative phase ($\phi_2-\phi_1\neq0,\pi$), giving rise to a complex $s+is'$ superconducting state with broken time-reversal symmetry. Combined with a nonsinusoidal Josephson current-phase relation, this state produces spontaneous nonreciprocal transport.}

\label{fig4} 
\end{figure*}

To disentangle the different contributions to the highly asymmetric pattern, we investigate the low-field response after zero-field cooling to minimize the influence of trapped vortices (Fig.~\ref{fig2}f). From the first Fraunhofer minimum at $B_0 \simeq$ 4.2\,mT, we estimate an effective junction width $W=\Phi_0/(B_0 d)\simeq 6 \mu m$, where $\Phi_0$ is the flux quantum and d$\sim$ 80\,nm is the total thickness of the heterostructure, in good agreement with the geometrical dimensions of the junction. 

The positive and negative switching currents, $I_{\rm sw}^{\pm}(B)$, obtained from the low-field measurements (Fig.~\ref{fig3}a), reveal nonreciprocal transport as quantified through the switching-current difference $\Delta_{\rm sw}=I_{\rm sw}^{+}-|I_{\rm sw}^{-}|$ and the diode efficiency $\eta=(I_{\rm sw}^{+}-|I_{\rm sw}^{-}|)/(I_{\rm sw}^{+}+|I_{\rm sw}^{-}|)$ (Fig.~\ref{fig3}b). Three notable features stand out. First, the diode response remains finite at zero applied magnetic field, whose microscopic origin is addressed later. Second, the switching currents reach maxima near $B=B^{\star}$, with $B^{\star}\simeq2$\,mT, and display opposite and approximately linear field dependences below $B^{\star}$. Third, above $B^{\star}$, both switching currents are rapidly suppressed and the diode polarity reverses.

The characteristic field $B^{\star}$, of the same order as the thermodynamic lower critical field reported for related TMD superconductors~\cite{kvashnin_coexistence_2020,Maj2020}, separates two distinct regimes of the nonreciprocal response. Below $B^{\star}$, the quasi-linear field response is phenomenologically described by $I_{c}^{\pm}= I_{c} \mp aB$, consistent with nonreciprocity arising from screening currents and asymmetric edge critical currents~\cite{hou23}. Above $B^{\star}$, the Josephson response is calculated using a two-harmonic current-phase relation and the phase distribution generated by the orbital phase gradient and a vortex-induced phase contribution. The observed polarity reversal is reproduced only when both a non-sinusoidal current-phase relation and an asymmetric phase distribution are included (Fig.~\ref{fig3}c and Extended Data Fig.\,3), identifying these as the minimal ingredients of the high-field nonreciprocal response.

However, these mechanisms cannot account for the finite nonreciprocal response at zero magnetic field. At high fields, the response is predominantly odd in B, consistent with conventional field-induced nonreciprocity and approximately satisfies the reciprocity relation $I_{\rm sw}^{+}(B)=-I_{\rm sw}^{-}(-B)$ (Fig.~\ref{fig3}d).  In contrast, a finite even-in-B component emerges at low fields and remains finite at nominal zero field, revealing a spontaneous contribution to the nonreciprocal response.
To exclude trapped vortices or remanent magnetic fields as the origin of the zero-field Josephson diode effect, we measured the nonreciprocal response after repeated thermal cycles using both zero-field-cooled and field-cooled protocols (Fig.~\ref{fig3}e). Within the experimental uncertainty determined by the switching-current distributions, neither the magnitude nor the sign of the diode effect changes appreciably, arguing against a flux-pinning origin.

To further exclude extrinsic mechanisms, we performed switching-current statistics at sub-Kelvin temperatures in the complete absence of any magnetic source (Fig.~\ref{fig3}f). The distributions reveal a robust zero-field Josephson diode effect over the temperature range investigated, with the expected broadening arising from thermally activated phase-slip processes. Similar behaviour is observed across multiple devices (Extended Data Fig.\,4), indicating that the zero-field Josephson diode arises from an intrinsic mechanism that spontaneously breaks time-reversal symmetry.

The hybrid molecular Josephson platform further provides insight into the microscopic origin of the zero-field nonreciprocal response. A spontaneous Josephson diode effect requires both an anomalous superconducting phase shift and a nonsinusoidal current-phase relation~\cite{mar23a,pal22,dav22}. To determine whether the anomalous phase originates at the interface or within the superconducting electrodes, we reversed their stacking order, forming a NbSe$_2$/TMA–TaS$_2$ heterostructure. The junction exhibits a superconducting transition at $T_{c}\simeq$ 2.6\,K (Fig.~\ref{fig4}a), governed by the intercalated TaS$_2$ electrode and comparable to that of the original device (Fig.~\ref{fig2}). Unlike the original configuration, the reversed junction shows no detectable zero-field diode response (Fig.~\ref{fig4}b). This comparison indicates that the anomalous phase originates within the superconducting state, while its manifestation in the Josephson diode response depends on the electrode configuration and transport geometry.

To probe the effect of molecular intercalation on the superconducting state, we subsequently intercalated the bottom NbSe$_2$ electrode with CTA$^+$ in the same device. Following intercalation, the junction retains coherent Josephson transport, while its transition temperature decreases to $T_{c}\simeq$ 0.9\,K (Fig.~\ref{fig4}c), as reported for monolayer NbSe$_2$~\cite{Wang2017}. Crucially, the emergence of the zero-field diode response after NbSe$_2$ intercalation (Fig.~\ref{fig4}d and Extended Data Fig.\,5) identifies molecular intercalation as a tuning knob for the internal phase structure of layered superconductors. The diode response is therefore predominantly sensitive to the superconducting state of the bottom electrode, likely owing to the asymmetric current paths imposed by the bottom-contact geometry (see Methods), together with the predominantly in-plane transport previously reported in molecularly intercalated TaS$_2$~\cite{Mar2025}. This interpretation is consistent with the field-free diode response observed in NbSe$_2$-based homojunctions~\cite{Ma2025}, which is restricted to the thin-flake regime.

These observations can be understood within a microscopic framework involving a time-reversal-symmetry-breaking multiband s+is\vtick\ superconducting state~\cite{yer24,mar26}.
In this scenario, interband scattering can drive competing real superconducting states into a complex multiband state characterized by a nontrivial relative phase.
%interband scattering between interacting superconducting condensates can drive competing real multiband superconducting states into a complex superconducting state characterized by a nontrivial relative phase. 
When combined with a nonsinusoidal Josephson current-phase relation, this internal phase structure generates spontaneous nonreciprocal transport (Fig.~\ref{fig4}e).
Supporting this picture, in the reduced-dimensionality regime accessed through molecular intercalation, interband scattering exceeding the superconducting gap scales has been independently identified spectroscopically in both TaS$_2$ and NbSe$_2$~\cite{Simon2026}. 
%The two TMDs also exhibit comparable superconducting energy scales in the 2D limit~\cite{sim24}, 
This suggests that the complex multiband scenario is not specific to a particular material, but may emerge more generally in multiband superconductors with substantial interband scattering.

In summary, we establish hybrid molecular Josephson junctions as a phase-sensitive platform to probe and engineer unconventional superconducting states in layered materials, and to integrate monolayer-like vdW superconductors into coherent quantum devices. 
%Our results reveal molecular intercalation as a means to control the internal superconducting phase structure and its manifestation in nonreciprocal Josephson transport.
The complex multiband scenario identified here provides a common framework for unconventional superconductivity reported in few-layer NbSe$_2$~\cite{Ma2025}, hybrid TaS$_2$ systems~\cite{24_chiral}, and related 1H/1T polymorphs~\cite{chiral4Hb,Liu2024,Sazgari2026G}, whose microscopic origin remains unresolved.
More broadly, this hybrid molecular platform enables the incorporation of molecular functionality into superconducting quantum devices. Extending this approach to photoactive, magnetic or switchable molecular systems, as well as to the broader family of transition-metal dichalcogenides, opens new opportunities to identify, realize and control emergent superconducting phases in layered quantum materials.\\

%\vspace{0.5cm}
\noindent\large{\textbf{Methods}}\normalsize\\
\small\textbf{Sample fabrication and structural characterization.} 
Heterostructure devices were fabricated in an Ar-filled glovebox using a polymer-based dry-transfer technique. Thin 2H-NbSe$_2$ and 2H-TaS$_2$ flakes were mechanically exfoliated onto Si/SiO$_2$ substrates and identified by optical contrast. The heterostructures were assembled using a deterministic pick-up method based on a polycarbonate (PC) film supported on a glass slide. The PC stamp was brought into contact with selected flakes at elevated temperature ($\sim$40--70$^\circ$C) to sequentially pick up and stack the layers. The assembled heterostructures were subsequently released onto pre-patterned Ti (2 nm)/Au (8 nm) electrodes on a Si/SiO$_2$ substrate. Finally, the PC film was dissolved in chloroform, yielding the completed van der Waals devices. No capping layer was employed, as the exposed vdW crystals were required for molecular intercalation.
Galvanic intercalation was performed under ambient conditions in less than 1\,h to minimize air exposure, following the procedures described in Refs.~\cite{Tezze2026,Mar2025}.
Selective tetramethylammonium (TMA) intercalation of the TaS$_2$ electrode was achieved using indium as the sacrificial anode, whose reduction potential lies between those of TaS$_2$ and NbSe$_2$. Under these conditions, TaS$_2$ intercalates preferentially while NbSe$_2$ remains unaffected. Control devices with intercalated NbSe$_2$ were subsequently obtained using magnesium as the anode, whose lower reduction potential enables intercalation of both materials.
The devices employ a bottom-contact geometry. Consequently, current injected through the lower flake acquires an out-of-plane transport component across the heterostructure, whereas transport within the upper electrode remains predominantly in-plane owing to the strong transport anisotropy of TMDs.
High-angle annular dark-field (HAADF) scanning transmission electron microscopy (STEM) imaging studies were performed in a double-aberration-corrected Thermo Fisher Scientific Titan Cube and in a probe-corrected Thermo Fisher Scientific Titan Low Base microscopes. The STEM energy dispersive X-ray spectroscopy (EDS) analyses were carried out using the double-aberration-corrected microscope. All these studies were conducted at 300 kV.
\\

\noindent\textbf{Transport measurements. }%\small
Electrical transport measurements were performed in a cryogen-free Quantum Design Physical Property Measurement System (PPMS) equipped with a sample rotator and in a $^3$He-$^4$He dilution refrigerator equipped with two-stage RC and $\pi$ filters, using a standard four-probe configuration. The PPMS was used for in-plane magnetic-field measurements, with the sample orientation adjusted to better than $0.2^\circ$. The dilution refrigerator was used for zero-field measurements and switching-current statistics, for which the superconducting magnet was physically removed to avoid residual or trapped magnetic flux.
The devices employ a bottom-contact geometry. Consequently, current injected through the lower flake acquires an out-of-plane transport component across the heterostructure, whereas transport within the upper electrode remains predominantly in-plane owing to the strong transport anisotropy of TMDs.
Current-voltage characteristics were acquired by sweeping a low-noise DC current bias in both directions while recording the voltage drop across the junction using a room-temperature low-noise preamplifier. Switching-current distributions and associated uncertainties were extracted from 1000 repeated current-voltage sweeps, acquired in two equally populated sets with reversed current-sweep sequences. $I_{sw}^{\pm}$ were determined using the half-normal-state voltage criterion. Joule heating was minimized by automatically returning the current to zero after the junction switched into the resistive state. Temperature stability was maintained within 20\,mK, with a delay introduced between successive switching measurements to ensure thermalization before the subsequent sweep.
No measurable variation of the nonreciprocal response was observed upon reversing the current-sweep sequence, introducing additional delay between sweeps, or repeating measurements over multiple cooling cycles. These observations further disfavour extrinsic contributions to the observed nonreciprocal response arising from the measurement protocol, instrumental asymmetry or thermal artefacts.\\

 \noindent\textbf{Numerical simulations. }%\small
The magnetic-field dependence of the nonreciprocal response was modelled by calculating the Josephson supercurrent $I=\sum_{n=1,2}\int_{-W/2}^{W/2} I_n \sin \left(n\phi(x) + \phi_0 \right) dx$, where
$\phi(x)=
2\pi \frac{B L_{\mathrm{eff}} x}{\Phi_0}
+
\varphi_v(x)$
contains both the conventional orbital phase gradient induced by the applied magnetic field and an additional vortex-induced phase texture. Here, $L_{\mathrm{eff}}$ denotes the effective length entering the orbital phase gradient. The latter is described by $\varphi_v(x)=\gamma_v  \arctan \frac{x-x_0}{z_0}$, where $\gamma_v$ controls the sign and magnitude of the phase winding, while ($x_0,z_0$) denote the lateral position and distance of the vortex from the junction, respectively, expressed in units of the junction width W. 
The critical currents were obtained by maximizing the supercurrent with respect to the integration constant $\phi_0$.
Below the characteristic field $B^{\star}$, the nonreciprocity was described phenomenologically by  $I_c^{\pm} = I_c \mp aB$ following Hou \emph{et al.}~\cite{hou23}. Above $B^{\star}$, the Josephson response was calculated using the phase distribution $\phi(x)$, including both the orbital phase gradient and the vortex-induced phase contribution. 
The model parameters ($I_2/I_1$, $\gamma_v$, $x_0$, and $z_0$) were varied to reproduce qualitatively the experimental Fraunhofer pattern and field-dependent nonreciprocal response. While $I_{2}/I_{1}$ and $x_{0}$ govern the sign reversal, the vortex distance from the junction primarily controls the magnitude of the nonreciprocal response. The complete model description together with the calculated Fraunhofer patterns, switching-current differences and diode efficiencies, is presented in Extended Data Fig.\,3.\\

\noindent\large{\textbf{Data availability}}\small\\
\noindent The data that support the findings of this study are available
from the corresponding author, Daniel Margineda, upon reasonable request.\\

\noindent\large{\textbf{References}}\normalsize
\vspace{-0.5cm}
\bibliography{bib_hybJJs}

\vspace{1cm}

\noindent\large{\textbf{Acknowledgments}}\normalsize\\
\small
\noindent This work was supported by MICIU/AEI/10.13039/501100011033 (Grant CEX2020-001038-M), and by MICIU/AEI and ERDF/EU (Projects PID2024-157558NB-C22, PID2024-155708OB-I00, and PID2023-148225NB-C31). M.G. acknowledges support from MICIU/AEI and EU NextGenerationEU/PRTR (Fellowship RYC2021-031705-I). C.A.-G. acknowledges support from MICIU/AEI and ESF+ (Fellowship PRE2022-104487).
This work was also funded by MICIU/AEI and by the European Union NextGeneration EU Plan (PRTR-C17.I1), by the IKUR-Quantum Technologies Strategy under the collaboration agreement between DIPC, CFM, and CIC nanoGUNE on behalf of the Department of Education of the Basque Government. R.A acknowledges funding from the Spanish MICIU with funding from the Spanish MICIU (PID2023-151080NB-I00/AEI/10.13039/501100011033 and CEX2023-001286-S MICIU/AEI /10.13039/501100011033), as well as from the Government of Aragon (DGA) through the project E13 23R. The TEM studies were conducted at the Laboratorio de Microscopías Avanzadas (LMA), Universidad de Zaragoza, Spain.\\

\noindent \large{\textbf{Author contributions}}\normalsize\\
\small \noindent D.M. fabricated the samples, performed the experiments, analysed the data. D.M and M.G wrote the manuscript with inputs from all authors. D.C. and M.I. provided experimental support and optimized the acquisition systems of the low-temperature cryostat. C.A. and D.T. optimized the intercalation method.  R.A. carried out TEM experiments. M,G., C.R. and L.H  coordinated the project. All authors discussed the results and their implications equally at all stages.\\

\noindent\large{\textbf{Additional Information}}\small\\
\textbf{Supplementary Information} The online version contains supplementary material. \\

\noindent\textbf{Correspondence} and requests for materials should be addressed to Daniel Margineda and Marco Gobbi

\onecolumngrid
\newpage
\section{Supplementary Information of  ``Molecule-intercalated van der Waals Josephson junctions uncover complex multiband superconductivity"}

\setcounter{figure}{0}
\renewcommand{\figurename}{\textbf{
Extended Data Fig.}}

\begin{figure*}[ht]
\centering
\includegraphics[scale=0.26 ]{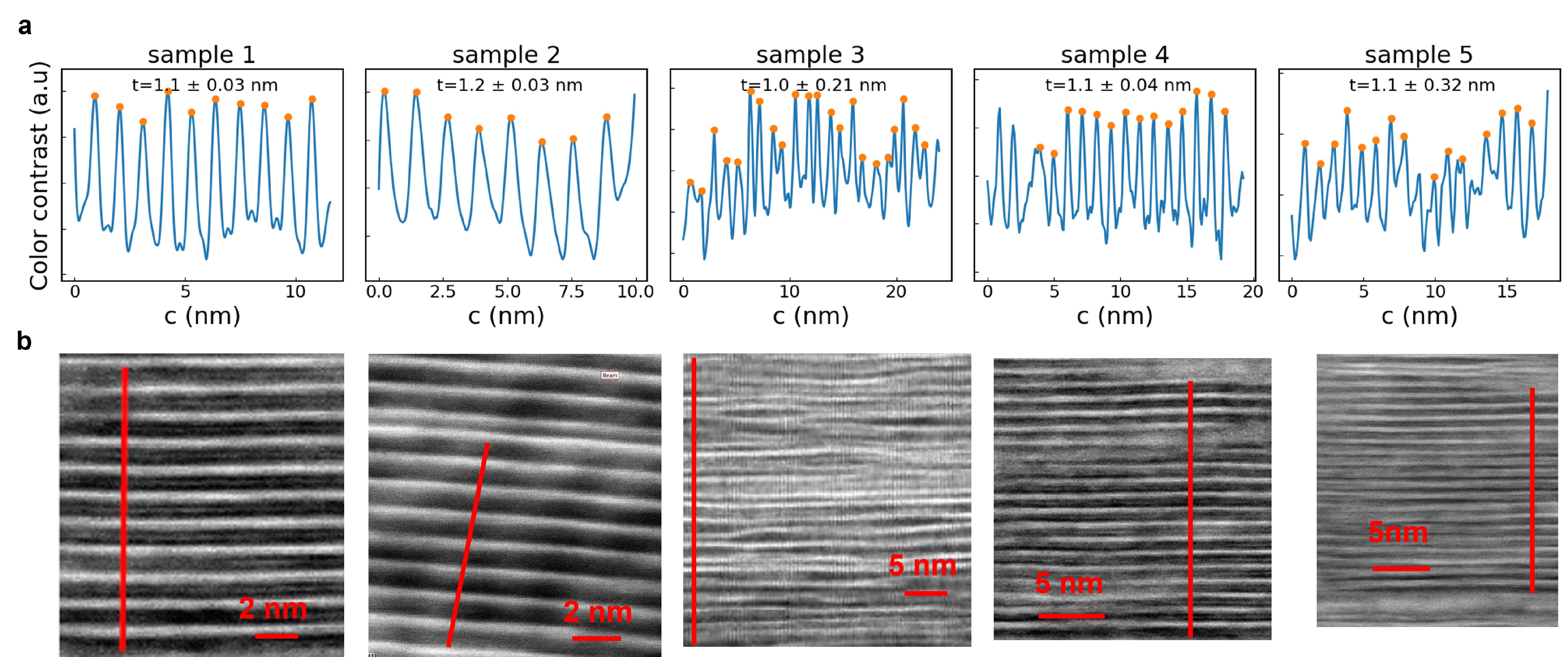}

\caption{
\textbf{Structural characterization of molecularly intercalated TaS$_2$.}
\textbf{a,} Line profiles extracted from high-resolution scanning transmission electron microscopy (HRSTEM) images of five representative intercalated crystals. All samples exhibit a uniform average interlayer periodicity consistent with homogeneous molecular expansion of the van der Waals gap.
\textbf{b,} Representative cross-sectional HRSTEM-HAADF images of the corresponding crystals. Red lines indicate the directions used to extract the line profiles shown in \textbf{a}, confirming highly ordered layered stacking over extended regions. Occasional local stacking faults are observed in some crystals, contributing to the dispersion of the measured interlayer spacing.
}

\label{figSI1}
\end{figure*}

\begin{figure*}[ht]
\centering
\includegraphics[scale=0.26 ]{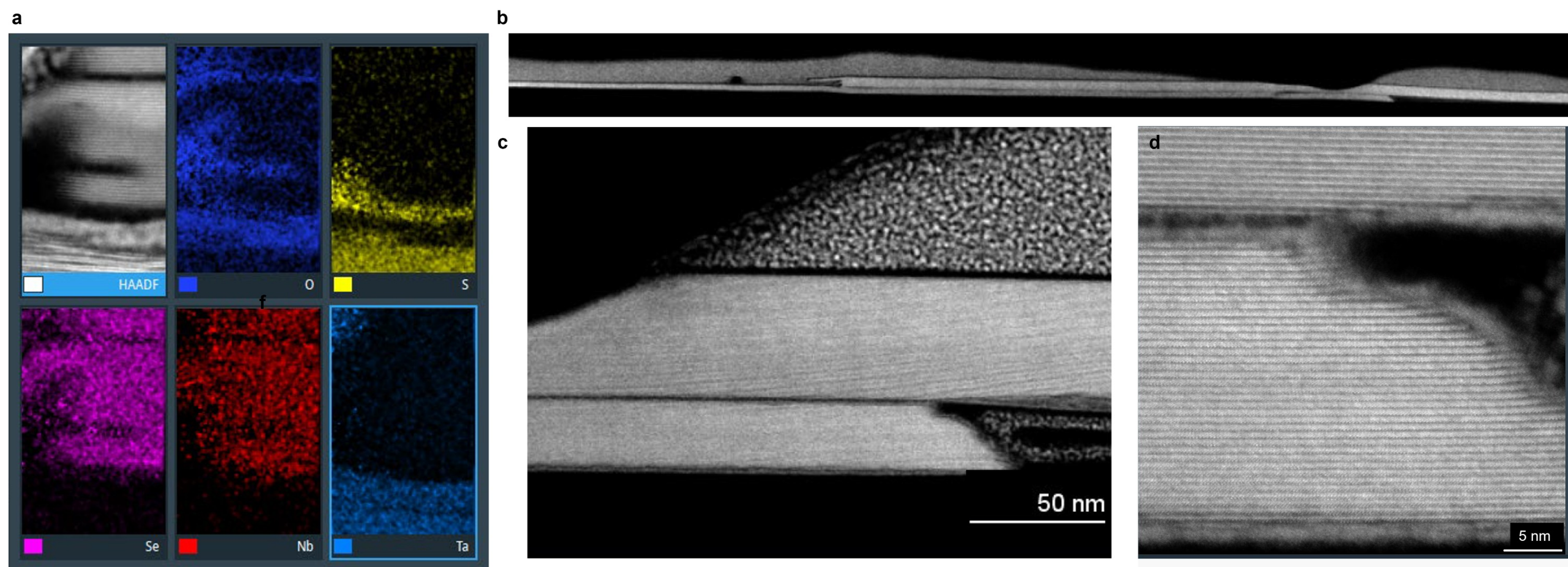}
\caption{
\textbf{HAADF-STEM characterization of TaS$_2$/NbSe$_2$ heterostructures.}
\textbf{a,} High-angle annular dark-field scanning transmission electron microscopy (HAADF-STEM) image of the interface in a molecularly intercalated TaS$_2$/NbSe$_2$ heterostructure together with the corresponding energy-dispersive X-ray spectroscopy (EDS) elemental maps of O, S, Se, Nb and Ta.
\textbf{b,} Low-magnification cross-sectional HAADF-STEM image of an unintercalated TaS$_2$/NbSe$_2$ heterostructure.
\textbf{c,} Higher-magnification HAADF-STEM image of the heterostructure interface.
\textbf{d,} Atomic-resolution HAADF-STEM image revealing a thin Ta-oxide interfacial layer. The presence of the oxide in the unintercalated heterostructure indicates that oxide formation originates from device fabrication rather than from the molecular intercalation process.}

\label{figSI2}
\end{figure*}

\begin{figure*}[ht]
\centering
\includegraphics[scale=0.45 ]{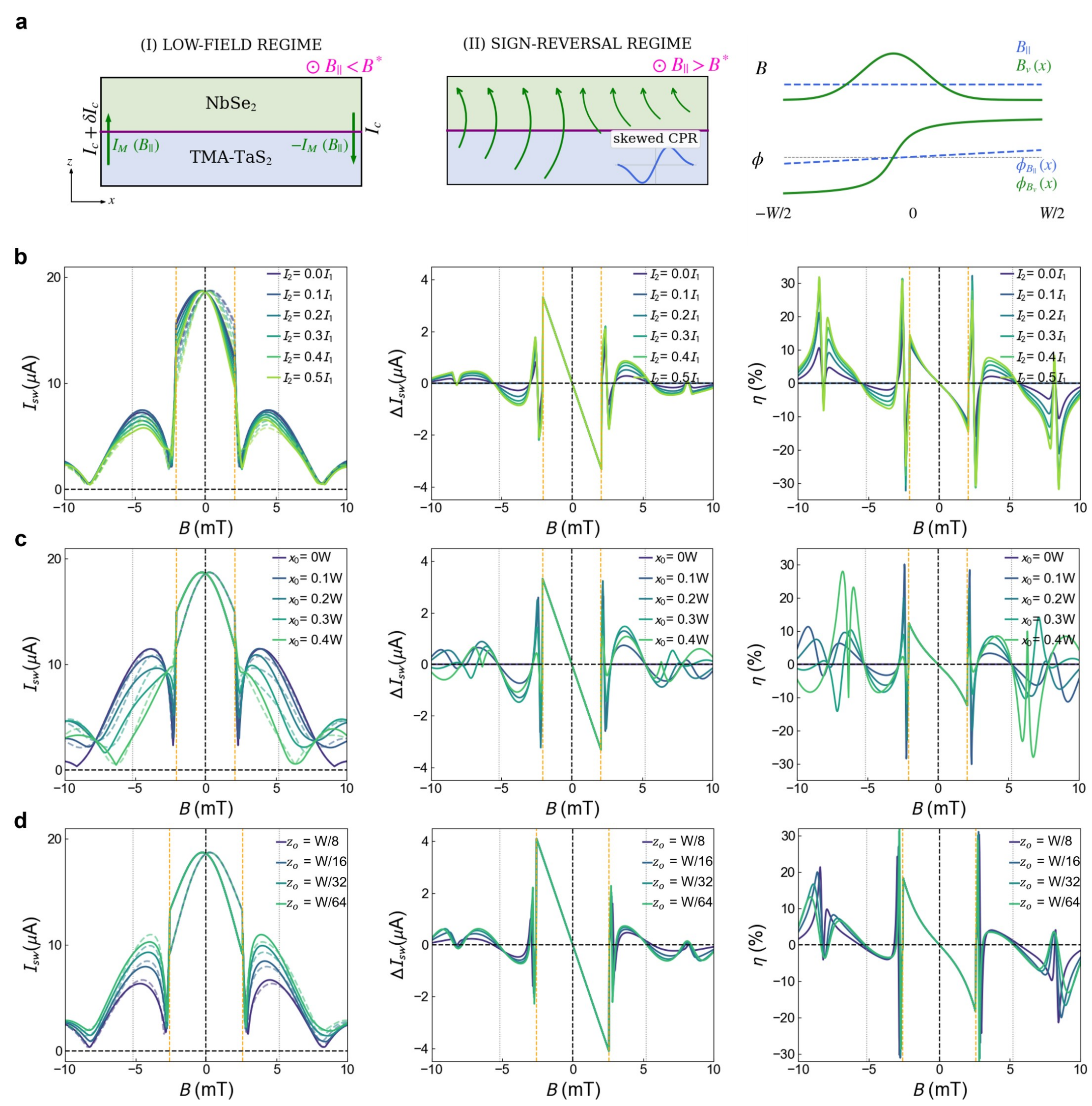}

\caption{\textbf{Phenomenological modelling of the field-dependent nonreciprocal response. a,} Below the characteristic field ($B<B^\star$), the low-field response is illustrated schematically by opposite screening currents and asymmetric edge critical currents, yielding the linear field dependence  $I_{c}^{\pm }=I_{c}\mp aB$. Above the characteristic field ($B>B^\star$), the nonreciprocal response is modelled using an orbital phase gradient and a vortex-induced phase texture, together with a skewed current-phase relation.  Representative magnetic-field and phase distributions across the junction width used in the high-field model are illustrated. \textbf{b--d,} Calculated diode efficiency as a function of magnetic field for different amplitudes of the second harmonic (\textbf{b}), and different vortex positions $(x_0,z_0)$ (\textbf{c,d}). The model captures the key experimental features, indicating that both a finite second harmonic and an asymmetric vortex-induced phase texture are necessary to reproduce the experimentally observed field-induced nonreciprocal response.}

\label{figSI3}
\end{figure*}

\begin{figure*}[ht]
\centering
\includegraphics[scale=0.27 ]{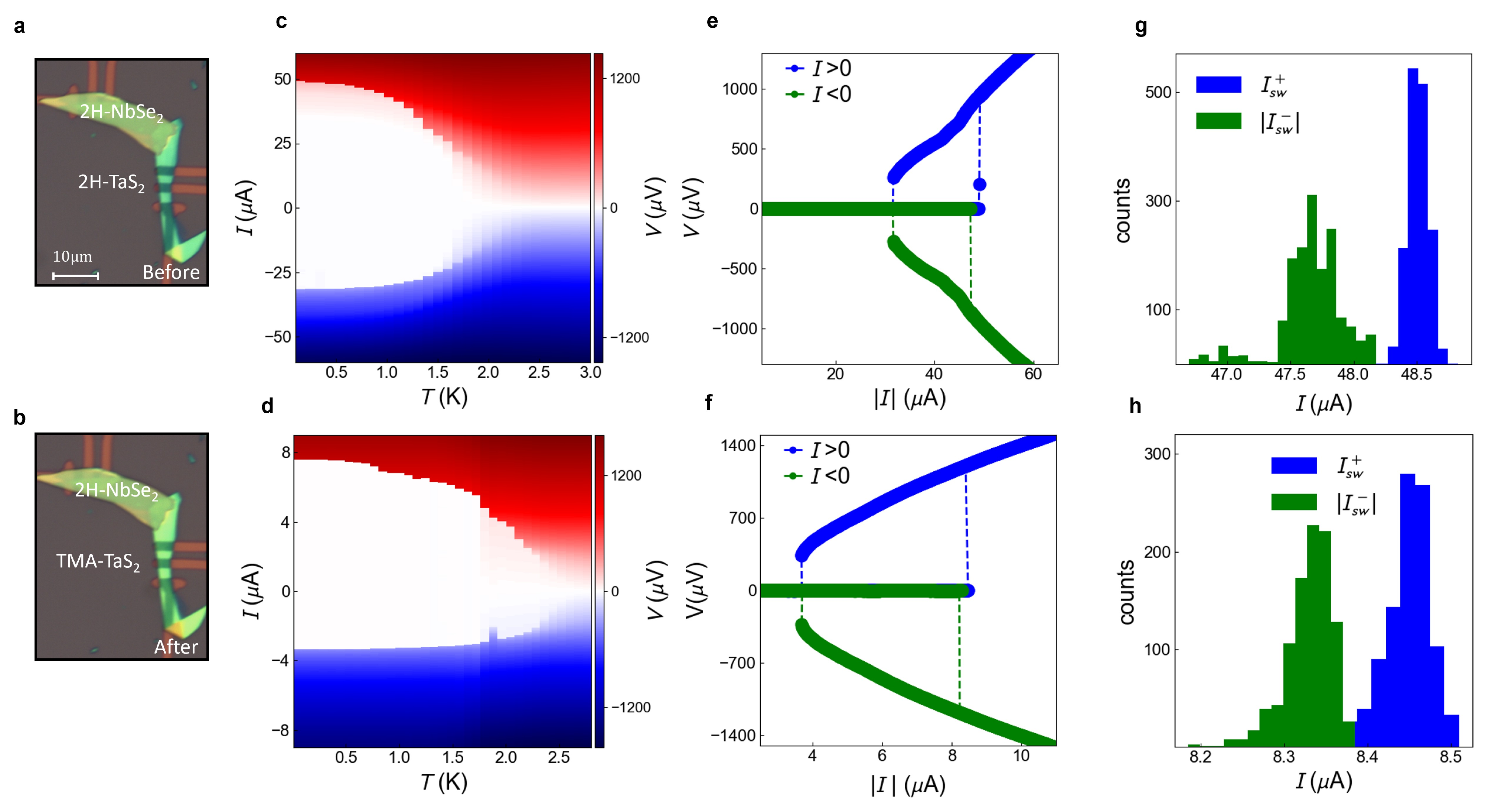}

\caption{
\textbf{Reproducibility of the spontaneous Josephson diode effect.}
\textbf{a,b} Optical micrographs of a TaS$_2$/NbSe$_2$ heterostructure before and after selective intercalation of the  TaS$_2$ electrode.
\textbf{c,d,} Temperature dependence of the current-voltage characteristics for two independent Josephson junctions, showing superconducting transition dominated by the intercalated TaS$_2$ electrode.
\textbf{e,f,} Representative current-voltage characteristics measured at 100\,mK for the devices shown in \textbf{c,d}, respectively.
\textbf{d,e} Switching-current distributions extracted from repeated current-voltage measurements , demonstrating reproducible spontaneous zero-field Josephson diode effects in both devices.}
\label{figSI4}
\end{figure*}

\begin{figure*}[ht]
\centering
\includegraphics[scale=0.27 ]{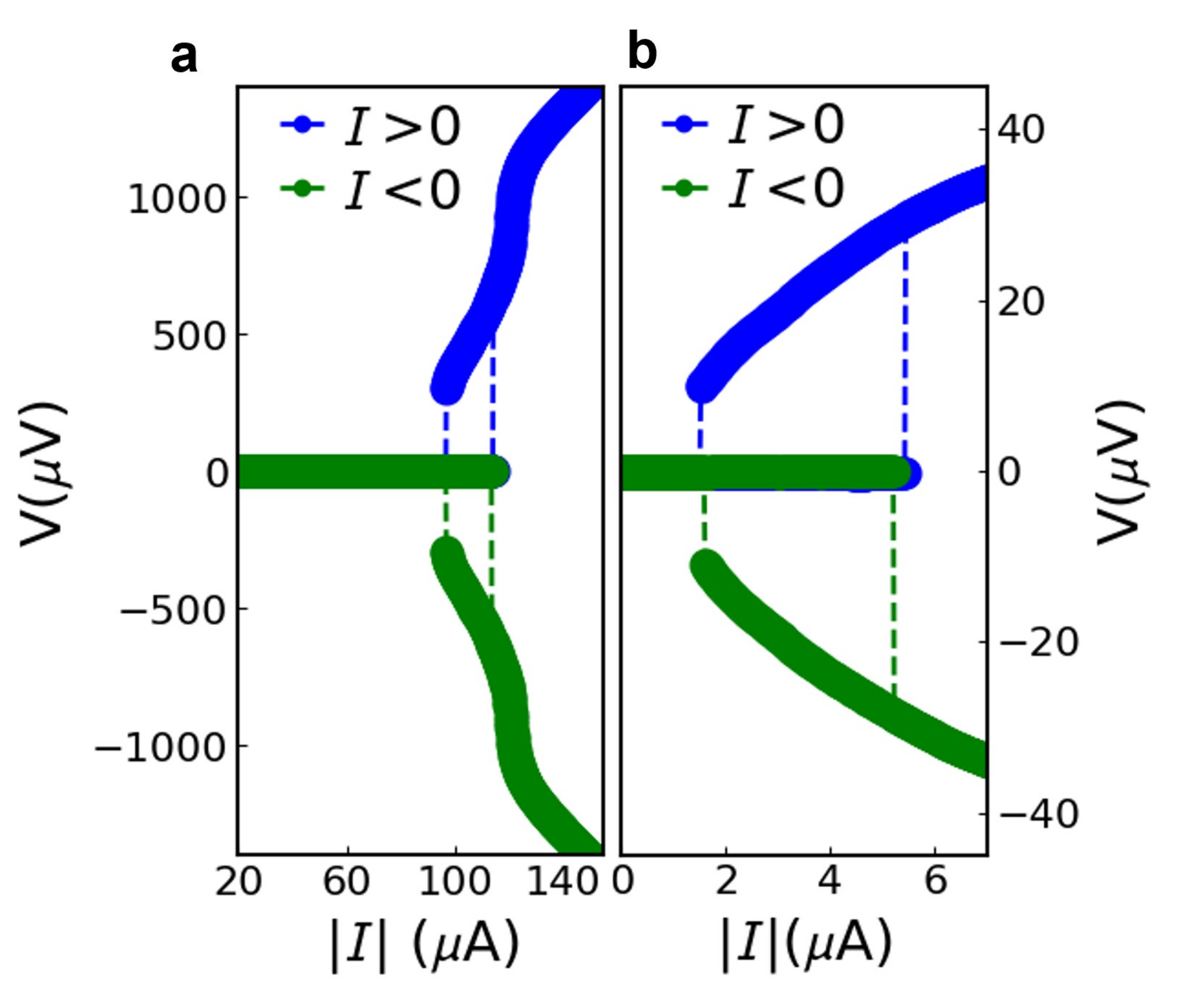}
\caption{\textbf{Superconducting diode effect.}\textbf{a,b} Representative current-voltage characteristics measured at 100\,mK for the devices shown in Fig.\,4 of the main text before and after NbSe$_2$ intercalation, respectively.}
\label{figSI4F}
\end{figure*}

\end{document}